\documentclass[12pt]{article}
\begin{document}
\hspace{4 in} USM-TH-146
\begin{center}  {Review of Spectroscopic Determination of Extra Spatial Dimensions in the Early Universe}
\end{center}
\vspace{.2in}
\begin{center}
Douglas J. Buettner
\\Aerospace Corporation\\P.O. Box 92957  - M1/112\\
Los Angeles, CA  90009-2957
\footnote{e-mail address:
Douglas.J.Buettner@aero.org}
\end{center}
\vspace{.2in}
\begin{center}
P.D. Morley
\\General Dynamics Corporation\\National Systems Group\\14700 Lee Road
Chantilly, VA 20151\footnote{e-mail address:
peter.morley@gd-ais.com}
\end{center}
\vspace{.2in}
\begin{center}
Ivan Schmidt
\\Department of Physics \\ Universidad T\'{e}cnica Federico
Santa Mar\'\i a \\ Casilla 110-V, Valpara\'\i so, Chile\footnote{e-mail address:
ischmidt@fis.utfsm.cl}
\end{center}
\vspace{.2in}
\begin{abstract}
\baselineskip=0.7cm

Atomic and molecular transitions for high Z objects in the early
universe give bounds to the possible existence of extra spatial
dimensions $D = 3+\epsilon$. We review the theory and present
observational data, based on Lyman and Balmer hydrogen transitions of distant quasars
and CO rotational transitions of far away giant molecular clouds.
\vspace{.4in}
\end{abstract}

The work of I. S. is supported in part by Fondecyt (Chile) grant
1030355.

\newpage
\parskip=9pt

\section{Introduction}

We live in a four dimensional space-time world. This has been checked
experimentally with great precision [1]. Nevertheless, in
principle there does not seem to be a reason for this, and in fact
the universe might have any number of dimensions.

The physics of extra dimensions began with the work of Kaluza and
Klein. They proposed uniting Maxwell's theory of electromagnetism
and Einstein's theory of gravitation by embedding them into a
generally covariant five-dimensional space-time, whose fifth
dimension was curled up into a tiny ring which was not
experimentally observable. More complicated non-abelian theories
can be obtained in much the same way, by starting with more
dimensions and compactifying them in various ways.

In recent years the idea of extra dimensions has been resurrected.
The main reason is that the leading candidate for providing a
framework in which to build a theory which unifies all
interactions, superstrings, has been found to be mathematically
consistent only if there are six or seven extra spatial
dimensions. Otherwise the theory is anomalous. In conventional
string theory the compactification of these extra dimensions
occurs at very high scales, close to the Planck scale. Here the extra dimensions manifest themselves
mainly through threshold effects of heavy states with masses close
to the Planck mass. These models contain the Standard Model (SM)
of particle interactions gauge group, and can also accommodate the
Minimal Supersymmetric Standard Model. Their main interesting
feature is that there is an automatic unification of gauge
couplings at a scale $M_{un}$. Nevertheless, there is still a
hierarchy problem because we have no way of explaining why the
scales of particle physics are so different from those of gravity.

In the last years there have been several attempts to solve the
hierarchy problem by using extra dimensions [2]. If space-time is
fundamentally $(4+n)$-dimensional, the $4$-dimensional Planck mass
$M_{Pl}^{(4)} \simeq 1.22 \times 10 ^{19} GeV$ depends on the
$(4+n)$- dimensional Planck mass $M_{Pl}$ and the volume $V_c$ of
the compact extra dimensions through

$$M_{Pl}^{(4)^2} = M_{Pl}^{n+2} V_c.$$

Since no extra dimensions have been detected experimentally, the
compactification scale ($\sim 1/V_c^{1/n}$) would have to be much
smaller than the weak scale $(100 GeV)^{-1}$, and the particles and forces of the
SM (except for gravity) must be confined to the four dimensional
world volume of a three-brane. By taking $V_c$ large enough it is
possible to eliminate the hierarchy problem between the weak and
Planck scales.

In all the above scenarios it is essential to know as precise as
possible the number of space dimensions. Since at present this
number is three [1], and since in the very early universe there exists
the possibility of a larger number, then this indicates that it should
be possible to define an effective number of space dimensions
which changes continuously with time, from its very early universe
value to the present one. Moreover, by looking at relics of the
early universe, such as the cosmic microwave background or the
light emitted by quasars with a very high redshift, it should be
possible to measure deviations of the number of spatial dimensions, $D$, 
from its present (epoch) value of 3. This review looks at the spectroscopic method of determining 
$\epsilon = D -3$ in the early universe.

\section{Quantum Mechanical Energy Levels in D-dimensions}

The spacing of atomic and molecular energy levels varies with the
dimension of space. An intuitive way to understand this is the
following. Both the kinetic energy term and the Coulomb potential
in the Schr\"{o}dinger equation are related to the Laplacian
operator. This operator has the property of measuring the
difference between the average value of the field in the immediate
neighborhood of a point and the precise value of the field at the
point. How the field spreads out from the point depends on the
dimension of space it resides in. In particular, $\epsilon$
differences between two fractal geometries give rise to
corresponding $\epsilon$ differences in their Laplacians.

Under some circumstances, it is possible to solve the Schr\"{o}dinger equation in $D =
3+\epsilon$ dimension of space ($4 + \epsilon$ dimension of
space-time), using a Taylor expansion [3] about $D = 3$

\begin{equation}
<|E|>|_{D=3+ \epsilon} =<|E|>|_{D=3} + \frac{d <|E|> }{d
D}|_{D=3}\epsilon + \cdots  \; . \end{equation} 
The Hellmann-Feynman theorem is

\begin{equation}
\frac{d <|E|> }{d D}|_{D=3}=<| \frac{\partial H}{\partial
D}|_{D=3} |>
\end{equation}
where $H$ is the D-dimensional Hamiltonian. In the succeeding
sections, we will give the quantum mechanical hydrogen
transitions and the linear molecular rotational
energies in D-dimensions, but first we turn to the discussion of the technique needed to measure $\epsilon$.

\section{How to Determine $\epsilon$}
Using ancient light, we are faced with the problem of disentangling the cosmological redshift to obtain the light's true rest-frame wavelength, $\lambda_{epoch}$, associated with the earlier time-epoch. In order to do this, we use the fact that all light from the same astronomical object has the same cosmological $Z_{c}$:

\begin{equation}
        Z_{c} = (\lambda_{observed} - \lambda_{epoch}) / \lambda_{epoch} \; .
\end{equation}
In this equation, we are trying to deduce $\lambda_{epoch}$ to see if $\lambda_{epoch}\neq \lambda_{present}$. Clearly, an observation of a single line, even if that line can be identified (e.g. Ly$_{\alpha}$), cannot determine $\lambda_{epoch}$ because the value of $Z_{c}$ is unknown. The problem is solvable if we have {\em two} lines from the same object, which then allows us to uniquely determine $\lambda_{epoch}$, and thus $\epsilon$:

\begin{equation}
\epsilon = \frac{\tau_{M} \lambda_{1} - \lambda_{0}}
          {a - b \tau_{M}}
\end{equation}
where 

  \begin{equation}
   \tau_{M} = \frac{\lambda_{0M}}{\lambda_{1M}}
\end{equation}
with
$\lambda_{0M}$, $\lambda_{1M}$ the two measured redshifted lines, which have early universe epoch rest-frame
wavelengths $\lambda_{0} + a\epsilon$, $\lambda_{1}+b\epsilon$,
where $\lambda_{0}$, $\lambda_{1}$ are the present epoch (i.e.
laboratory) transition wavelengths. In this equation, $a$ and $b$ are determined by solving the Schr\"{o}dinger equation in $D = 3+\epsilon$ dimensions. 

Thus the game is to find observational data of high quality that has at least two identifiable transitions from the same deep-space emitter. 

\section{Rotational Energy Levels in D-dimensions}

We are interested in linear molecules, such as CO, because, being
the simplest molecules, they are the ones most likely to be
identified in distant, giant molecular clouds. Also, it turns out that the Schr\"{o}dinger equation in $D = 3+\epsilon$ dimensions is solvable in this case. 

The Hamiltonian is

\begin{equation} H_{rot} = B(L)L^{2}
\end{equation}
where $L$ is the body (molecule)-fixed rotational angular momentum
and $B$ is the principal rotational constant (equal to 1/2I, where
I is the principal moment of inertia). Because of centrifugal
stretching, $B = B(L)$. We need to generalize this energy to $D =
3 + \epsilon$ fractal space. In quantum mechanics, $L^{2}$ is a
second order Casimir invariant operator, $C_2$. Thus the generalized rotational operator is

   \begin{equation} H_{rot} = B(C_{2})C_{2}
\end{equation}
The piece $B(C_{2})$ has a simple form when no vibrations are excited [4]

  \begin{equation} B(C_{2}) = B_{0} + B_{1} C_{2} \; .
\end{equation}
In general, the second
order Casimir invariant is [5]:

\begin{equation}
C_2 = f^i_{jk} f^j_{il} X^k X^l = H_i G_{ij} H_j + \sum_{\rm all\
roots} E^{\alpha} E_{-\alpha}
\end{equation}
where $f^i_{jk}$ are the structure constants, $X_k$ are
generators, and $C_2$ commutes with all generators. The Racah
formula for the eigenvalue of $C_2$ for any irreducible
representation is easily  derived by letting $C_2$ act on the
state with highest weight $\Lambda$. The result is:

\begin{equation}
C_2 = (\Lambda,\Lambda+2\delta)
\end{equation}
where $\delta=(1,1,....,1)$ in the Dynkin basis. The scalar
product of any two weights can be written as:

\begin{equation}
(\Lambda,\Lambda') = \sum_{ij} a'_j G_{ij} a _j 
\end{equation}
where $G_{ij}$ is a symmetric tensor whose elements can be
computed  for each simple group, and which are given in Table 7 of [5],
and the $a_i$ are the Dynkin components of $\Lambda$.

In our case we want to find the Casimir invariant for the $(L,0,
...0)$ totally symmetric representation. The choice of
representation depends, of course, on the way in which we want to
generalize angular momentum, and the correct choice, we argue,
would preserve the symmetry properties (in this case this would
mean to keep the completely symmetric coupling) when generalizing
to larger dimensions.

For odd space dimensions $D=2n+1, n=1,2,...$, the algebra is
$B_n$, and for even space dimensions $D  = 2n, n=1,2,...$, it is
$D_n$. We calculate $C_2$, using the equations above, and the $a_j$ values given by $(L,0, ...0)$:

\begin{equation}
C_2 = (L,0,....,0) G(B_n\ {\rm or}\ D_n) (2+L,2,2,....,2)
\end{equation}
which gives, by simple matrix multiplication, both for $B_n$ and
$D_n$:

\begin{equation}
C_2 = L (L+D-2) \  .
\end{equation}
The pure (no vibrational quanta) rotational energies in $D$-dimension space are then:

\begin{equation}
{\cal H}_{rot} = [B_{0} + B_{1}(L(L+1))]L(L+1) + \{2B_{1}L^{2}(L+1) + B_{0}L \} \epsilon \; .
\end{equation}   

\section{Hydrogen Lyman Transitions}

The hydrogen atom is the only other system that is amendable to solution in $D = 3 + \epsilon$ dimensions.
By means of the Taylor expansion, we obtain in Table 1 the Lyman
transitions as a function of $\epsilon$, for $\epsilon \ll 1$. If
$\epsilon \neq 0$, the result is a change in each transition by a
unique amount. The effect is unmistakable: even a small shift
$\epsilon \sim 0.03$ will be detectable.

Lyman data has been analyzed in reference [3]. Except for anomalous data associated with one data set @ $Z_{c} \sim 4$, $<\epsilon> \approx 0$.
High $Z_{c}$ quasars are extremely rare, but the Sloan Digital Sky
Survey (http://www.sdss.org) has discovered several $Z_{c} \geq
5$. Looking at each Sloan spectrogram, one has difficulty in
identifying the original center-line of the Lyman series. The
spectrum is a superposition of a priori unknown set of
cosmological red shifts. Even the largest or primary shift
introduces a complicated transposed spectrum. Next, uneven
absorption of radiation as it moves through one intergalactic
cloud to another can remove completely the center of a spectral
line or one of its wings and may lead to complete masking of the
location of the original radiated line. Lastly, line overlaps,
which are such a rare occurrence in terrestrial plasmas is the
norm for high $Z_{c}$ spectra. These circumstances have lead to
the observation that no astronomy group has been able to locate
the matching (same cloud or put in another way, identical $Z_{c}$
values) Lyman series Ly$_{\alpha}$ to Ly$_{\epsilon}$ in one
$Z_{c} \geq 4$ spectrogram. Due to absorption, the Lyman series in
hydrogen becomes difficult to use for $<\epsilon>$ determination
for very high $Z_{c}\geq 4$.

     \begin{table}
     \begin{center}
     \begin{tabular}{cc}
     \multicolumn{2}{c}{$\epsilon$ Transitions} \\  \hline
     Lyman line  & wavelength  \\
     \mbox{} & \mbox{} \\
     Ly$_{\alpha}$  &  1215.67 + $1418.27\epsilon$ \\
     Ly$_{\beta}$  &  1025.72 + $1111.18\epsilon$ \\
     Ly$_{\gamma}$ &  972.537 + 1021.155$\epsilon$ \\
     Ly$_{\delta}$ &  949.743 + $981.391\epsilon$ \\
     Ly$_{\epsilon}$ & 937.803 + $960.122\epsilon$
     \end{tabular}
     \end{center}
     \caption{$\epsilon$ Lyman hydrogen quantum mechanical wavelengths
     (\AA ngstroms)}
     \end{table}

\section{Hydrogen Balmer Transitions}
The Balmer lines start off in the optical (for small $\epsilon \ll
1$) and get redshifted to the infrared for high $Z_{c}$. By using
the atmospheric infrared windows where absorption is small, the
infrared Balmer lines can evade the difficulties that potentially
plague the Lyman series. In Table 2, we give the hydrogen Balmer
epsilon-dependent transitions. 
The measured redshift of $(Z_{c}+1)\lambda_{0}$ means, for
example, a $Z_{c}=3$ quasar has Balmer lines in the infrared.

Various online databases were searched for papers containing
emission spectra from high Z$_{c}$ quasars that contained coincident
hydrogen Balmer lines. Of the many references identified in these
databases, only one had tabular data [6].
Another, [7], had a single Balmer line pair
measured. In the future, infrared data will be more plentiful as
various observatories bring on-line sophisticated infrared
spectrometers. We use eq(4) above to determine $\epsilon$. The uncertainty in $\epsilon$, $\delta \epsilon$, is determined by the equation in reference [3].

     \begin{table}
     \begin{center}
     \begin{tabular}{cc}
     \multicolumn{2}{c}{$\epsilon$ Transitions} \\  \hline
     Balmer line  & wavelength  \\
     \mbox{} & \mbox{} \\
     H$_{\alpha}$  &  6562.8 + $4155.24\epsilon$ \\
     H$_{\beta}$  &  4861.36 + $2834.99\epsilon$ \\
     H$_{\gamma}$ &  4340.49 + $2417.58\epsilon$
     \end{tabular}
     \end{center}
     \caption{$\epsilon$ Balmer hydrogen quantum mechanical wavelengths
     (\AA ngstroms)}
     \end{table}

     \begin{table}
     \begin{center}
     \begin{tabular}{cccccc}
     \multicolumn{6}{c}{Balmer $H_{\alpha}$ and $H_{\beta}$ } \\  \hline
     Quasar  & transition & wavelength (\AA ) & FWHM (km/s) &
     FWHM (\AA ) & $\sigma$            \\
     \mbox{} & \mbox{} & \mbox{} & \mbox{} & \mbox{} & \mbox{} \\
     Q0007-000 & $H_{\alpha}$ & 21572.1 & 4500 &   323.8 &  116.6 \\
      \mbox{} & $H_{\beta}$ &  15896.5 & 3500 &   185.6 &  66.9  \\
      Q0027+018 & $H_{\alpha}$  &    21952.7 & 4500  & 329.5 &  118.7 \\
       \mbox{} &  $H_{\beta}$ & 16091 &  4500 &   241.5 &  87.0 \\
      Q0237-233 & $H_{\alpha}$ &  21152.1 & 7500 &  529.2 &  190.6 \\
       \mbox{} & $H_{\beta}$ & 15702.1 & 3000 &  157.1 &  56.6 \\
       Q1623-268 & $H_{\alpha}$ & 23258.7 & 2600 &   201.7 &  72.7 \\
        \mbox{} &  $H_{\beta}$ & 17063.3 & 2500 &  142.3 &  51.3 \\
      Q1816+475 & $H_{\alpha}$ & 21132.4 & 2600 & 183.3 &  66.0 \\
      \mbox{} &  $H_{\beta}$ & 15653.5 & 2600 & 135.8 &  48.9
     \end{tabular}
     \end{center}
     \caption{Taken from reference [6]}
     \end{table}

     \begin{table}
     \begin{center}
     \begin{tabular}{cccc}
     \multicolumn{4}{c}{Balmer $H_{\alpha}$ and $H_{\beta}$ } \\  \hline
     Quasar  & transition & wavelength (\AA ngstroms) & $\sigma$ \\
     \mbox{} & \mbox{} & \mbox{} & \mbox{}  \\
     1937-101 & $H_{\alpha}$ &  23228.2 & 57.2 \\
      \mbox{} & $H_{\beta}$ & 20708.6 & 64.2
      \end{tabular}
       \end{center}
      \caption{Taken from reference [7]}
      \end{table}

It is seen that $<\epsilon> \simeq 0$ is favored by this small
sample of $Z_{c} \sim 3$ Balmer data. Balmer data for $Z_{c} \simeq 4$ would be
especially interesting.

     \begin{table}
     \begin{center}
     \begin{tabular}{ccc}
     \multicolumn{3}{c}{$\epsilon$ values} \\  \hline
       Quasar & $\epsilon$ & $\Delta\epsilon$ \\
       \mbox{} & \mbox{} & \mbox{} \\
       Q0007-000  &     0.11 &   0.15 \\
       Q0027+018  &     0.24 &   0.2  \\
       Q0237-233  &     -0.04 &  0.18  \\
       Q1623-268  &     0.22 &   0.12 \\
       Q1816+475  &     0   &    0.09 \\
       1937-101  &      0.059 &  0.16
      \end{tabular}
      \end{center}
      \caption{$\epsilon$ from Balmer spectra}
      \end{table}

\section{Molecular Rotational Transitions}

The Balmer quasar spectra allow a better determination of
$<\epsilon>$ than the Lyman data for high $Z_{c}$ objects, but the best spectroscopic
data would be lines starting off already in the microwave, and redshifted
towards the radio, namely rotational spectra.

The linear molecule $CO$ ($C^{12}O^{16}$) is the main non-hydrogen emitter in distant, giant molecular clouds. In Table 6, we give its (present epoch) laboratory transition frequencies [8]. It is a simple matter to determine 
the constants that appear in eq(8): $B_{0}=57.635968 \; {\rm GHz}$ and $B_{1}=-1.835 \times 10^{-4} \; {\rm GHz}$ for $CO$. By doing a literature search, we have identified 6 tabulations of $CO$ data that are usable, including data from the presently known farthest quasar J114816.64+525150.3 @ Z$_{c}$ = 6.42.

      \begin{table}
      \begin{center}
     \begin{tabular}{cc}
     \multicolumn{2}{c}{$C^{12}O^{16}$ Transitions} \\  \hline
     transition  & MHz  \\
     \mbox{} & \mbox{} \\
     $1 \rightarrow 0$  &  115271.202        \\
     $2 \rightarrow 1$  &  230538 \\
     $3 \rightarrow 2$ &  345795.991 \\
     $4 \rightarrow 3$ &  461040.77 \\
     $5 \rightarrow 4$  & 576267.922 \\
	$6 \rightarrow 5$   & 691473.09 \\
	$7 \rightarrow 6$  & 806651.719
     \end{tabular}
      \end{center}
     \caption{Laboratory CO pure rotational transitions}
     \end{table}

        \begin{table}
      \begin{center}
    \begin{tabular}{ccc}
     \multicolumn{3}{c}{Giant Molecular Clouds} \\  \hline
     identification  & observed transitions & source  \\
     \mbox{} & \mbox{} &  \mbox{} \\
     QSO J114816.64+525150.3 &  $7 \rightarrow 6, 6 \rightarrow 5, 3 \rightarrow 2$ & [9]      \\
	QSO H1413+117 &  $5 \rightarrow 4, 4 \rightarrow 3, 3 \rightarrow 2$ & [10] \\
    QSO PSS 2322+1944  & $5 \rightarrow 4, 4 \rightarrow 3, 2 \rightarrow 1, 1 \rightarrow 0$ & [11] \\
      QSO BR1202-0725  &  $7 \rightarrow 6, 5 \rightarrow 4$ & [12] \\
     QSO F10214+4724  & $6 \rightarrow 5, 3 \rightarrow 2$  & [12] \\
	QSO HR10 (J164502+4626.4)  & $5 \rightarrow 4, 2 \rightarrow 1$ & [13]
     \end{tabular}
      \end{center}
     \caption{Giant Molecular Cloud Transitions}
     \end{table}

     \begin{table}
      \begin{center}
    \begin{tabular}{cccccc}
     \multicolumn{6}{c}{Data} \\  \hline
     QSO & approx Z$_{c}$ &  transition & obs (Ghz) & FWHM & channel width  \\
     \mbox{} & \mbox{} & \mbox{} & \mbox{} & \mbox{}  & \mbox{} \\
     J114816.64+525150.3 & 6.42 &  $7 \rightarrow 6$ & 108.729 & 279 km/s & 5 MHz \\
	\mbox{} & \mbox{} & $6 \rightarrow 5$ & 93.204 &  279 km/s & 5 MHz \\
      \mbox{} & \mbox{} & $3 \rightarrow 2$ & 46.61 & 320 km/s &  50 MHz \\
      H1413+117  & 2.56 &  $5 \rightarrow 4$ & 161.964 & 398 km/s & 512 MHz \\
       \mbox{} & \mbox{}& $4 \rightarrow 3$ & 129.576 & 375 km/s & 512 MHz \\
	\mbox{} & \mbox{}& $3 \rightarrow 2$ & 97.199 & 362 km/s & 512 MHz \\
	PSS 2322+1944 & 4.12 & $5 \rightarrow 4$ & 112.55 & 273 km/s & 35 MHz \\
	\mbox{} & \mbox{}& $4 \rightarrow 3$ & 90.05 & 375 km/s & 35 MHz \\
	\mbox{} & \mbox{}& $2 \rightarrow 1$  & 45.035  & 200 km/s & 6.25 MHz \\
	\mbox{} & \mbox{}& $1 \rightarrow 0$ & 22.515 & 200 km/s & 50 MHz \\
	BR1202-0725 & 4.71 & $7 \rightarrow 6$ & 141.2 & 300 km/s & equivalent 60 km/s \\
	\mbox{} & \mbox{}& $5 \rightarrow 4$  & 101.3 & 350 km/s & equivalent 60 km/s \\
	F10214+4724 & 2.29 & $6 \rightarrow 5$  & 210.5 & 300 km/s & equivalent 80 km/s \\
	\mbox{} & \mbox{}&  $3 \rightarrow 2$  & 105.2 & 220 km/s & equivalent 80 km/s \\
	HR10 & 1.44 & $5 \rightarrow 4$ & 235.982 & 380 km/s & equivalent 75 km/s \\
	\mbox{} & \mbox{}& $2 \rightarrow 1$ & 94.405 & 400 km/s & equivalent 50 km/s   
     \end{tabular}
      \end{center}
     \caption{Early Universe CO cloud data}
     \end{table}

	\begin{table}
      \begin{center}
     \begin{tabular}{cccc}
     \multicolumn{4}{c}{Processed Data} \\  \hline
     QSO & line pairs & $\epsilon$ & $\Delta \epsilon$  \\
     \mbox{} & \mbox{} & \mbox{} & \mbox{} \\
	J114816.64+525150.3 & 7 $\rightarrow$ 6, 6 $\rightarrow$ 5  & -0.000012043  & 0.118937801 \\
	\mbox{} & 7 $\rightarrow$ 6, 3 $\rightarrow$ 2 & -0.000004379 & 0.084521824 \\
	\mbox{} &  6 $\rightarrow$ 5, 3 $\rightarrow$ 2 & -0.000003283 & 0.096631339 \\
	H1413+117 & 5 $\rightarrow$ 4, 4 $\rightarrow$ 3 & -8.1128E-04 &  7.2375E-01 \\
	\mbox{} & 5 $\rightarrow$ 4, 3 $\rightarrow$ 2 & 1.6508E-03 & 3.2882E-01 \\
	\mbox{} & 4 $\rightarrow$ 3, 3 $\rightarrow$ 2 & 3.1291E-03 & 5.6347E-01 \\
	PSS 2322+1944 & 5 $\rightarrow$ 4, 4 $\rightarrow$ 3 & 2.1489E-03 & 1.1919E-01 \\
	\mbox{} & 5 $\rightarrow$ 4, 2 $\rightarrow$ 1 & 1.3294E-03 & 1.3841E-02 \\
	\mbox{} & 5 $\rightarrow$ 4, 1 $\rightarrow$ 0 & 1.7313E-04 & 2.8838E-02 \\
	\mbox{} & 4 $\rightarrow$ 3, 2 $\rightarrow$ 1 & 1.1654E-03 & 2.0347E-02 \\
	\mbox{} & 4 $\rightarrow$ 3, 1 $\rightarrow$ 0 & 4.1410E-05 & 3.1010E-02 \\
	\mbox{} & 2 $\rightarrow$ 1, 1 $\rightarrow$ 0 & -5.2042E-04 & 4.5728E-02 \\
	BR1202-0725 & 7 $\rightarrow$ 6, 5 $\rightarrow$ 4 & 0.149826 & 0.0765425 \\
	F10214+4724 & 6 $\rightarrow$ 5, 3 $\rightarrow$ 2 & -0.00775587 & 0.0210779 \\
	HR10 & 5 $\rightarrow$ 4, 2 $\rightarrow$ 1 & -3.00361E-05 & 0.0132079
     \end{tabular}
      \end{center}
     \caption{$\epsilon$ from $CO$ spectra}
     \end{table}

\newpage

\section{Conclusion}
We take the high resolution data @ Z = 6.42 and perform the statistical Z-test [14] by 
taking $< \Delta \epsilon> $ as the standard deviation.  
This predicts that the probability of $\epsilon \neq 0$ is 1 in 7794, only 850 million years (using the standard cosmology) after the Big Bang.

The experimental spectroscopic data from ancient light shows that the dimension of space was 3 (present value) very soon after the Big Bang. The extra dimensions that some theories predict must either occur at very early time or somehow be restricted such that ordinary baryonic matter cannot couple to it.

\clearpage

\end{document}